\documentclass{article}
\usepackage{epsfig}

\newfont{\Bbb}{msbm10 scaled 1200} 
\newcommand{\mathbb}[1]{\mbox{\Bbb #1}}
\def\IC{{\mathbb C}}
\def\IR{{\mathbb R}}
\def\IZ{{\mathbb Z}}

\def\Poincare{{Poincar\'e }}

\def\TL{\hfil$\displaystyle{##}$}
\def\TR{$\displaystyle{{}##}$\hfil}

\def\lbldef#1#2{\expandafter\gdef\csname #1\endcsname {#2}}
\def\eqn#1#2{\lbldef{#1}{(\ref{#1})}%
\begin{equation} #2 \label{#1} \end{equation}}
\def\eqalign#1{\vcenter{\openup1\jot
 \halign{\strut\span\TL & \span\TR\cr #1 \cr
 }}}

\def\href#1#2{#2}

\def\sp{ \sin \phi}

\def\u1{{U(1)}}

\def\frac#1#2{{#1\over#2}}

\def\half{\frac12}

\def\d{\partial}

\def\inbar{\,\vrule height1.5ex width.4pt depth0pt}
\def\IC{\relax\hbox{$\inbar\kern-.3em{\rm C}$}}
\def\IR{\relax{\rm I\kern-.18em R}}
\def\IP{\relax{\rm I\kern-.18em P}}

\catcode`\@=11
\def\slash#1{\mathord{\mathpalette\c@ncel{#1}}}
\def\underrel#1\over#2{\mathrel{\mathop{\kern\z@#1}\limits_{#2}}}

\catcode`\@=12

\def\({\left(}
\def\){\right)}
\def\[{\left[}
\def\]{\right]}

\def \sinh{{\rm sinh}}
\def \cosh{{\rm cosh}}

\def\exp{{\rm exp}}
\def\sh{{\rm sinh}}
\def\ch{{\rm cosh}}

\usepackage{fortschritte}

\begin {document}

\hfill hep-th/0305140 \vskip .1in\hfill RI-05-03

\def\titleline{Beyond the Singularity of the
2-D Charged Black Hole}

\def\email_speaker{{\tt giveon@vms.huji.ac.il, eliezer@vms.huji.ac.il,
asever@phys.huji.ac.il }}
\def\authors{Amit Giveon, Eliezer Rabinovici, Amit Sever~\sp}

\def\addresses{
Racah Institute of Physics, The Hebrew University \\
Jerusalem, 91904, Israel.}


\def\abstracttext{Two dimensional charged black holes in string
theory can be obtained as exact ${SL(2,\IR)\times U(1)\over U(1)}$
quotient CFTs. The geometry of the quotient is induced from that
of the group, and in particular includes regions beyond the black
hole singularities. Moreover, wavefunctions in such black holes
are obtained from gauge invariant vertex operators in the
$SL(2,\IR)$ CFT, hence their behavior beyond the singularity is
determined. When the black hole is charged we find that the
wavefunctions are smooth at the singularities, and known
results of infinite blue shifts at the horizons persist. Unlike
the uncharged case, scattering waves prepared beyond the
singularity are not fully reflected; part of the wave is
transmitted through the singularity. Hence, the physics outside
the horizon of a {\it charged} black hole is sensitive to
conditions set behind the past singularity.}

\large

\makefront

\section{Introduction}

The two dimensional black hole in string theory \cite{efrmw,w,dvv}
is obtained as an exact $SL(2,\IR)\over U(1)$ quotient CFT
background \cite{w}. The geometry of this 2-$d$ black hole is
similar to a two dimensional slice of the Schwarzschild solution,
whose maximal analytic extension is shown in figure 1a in a
Kruskal diagram and its Penrose diagram is shown in Fig 1b. In the
4-$d$ case every point in figure 1 is actually a two sphere, while
in the 2-$d$ case there is a non-trivial dilaton.

 \begin{figure}[h]
 \begin{center} \epsfxsize = 4in
\epsffile{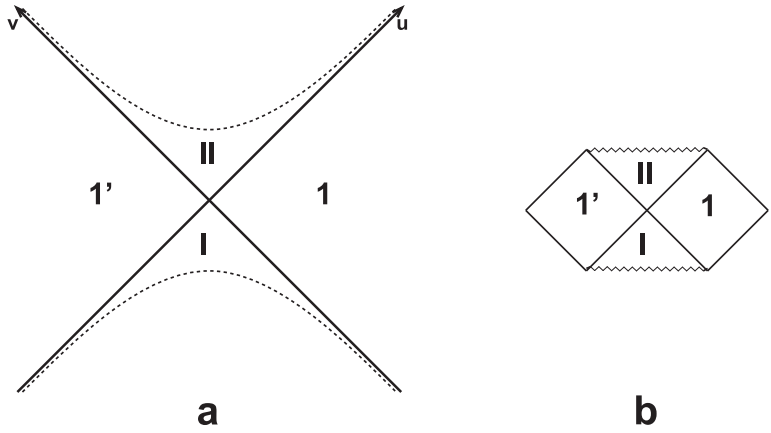} \caption{{\bf a}: Kruskal diagram and {\bf
b}: Penrose diagram of the Schwarzshild or the 2-$d$ black hole.}
\label{fig 1} \end{center}
 \end{figure}

In general relativity, regions beyond the singularities are not
considered usually, since initial data set outside the horizon of
the black hole cannot determine the behavior of classical
solutions beyond the singularities. In particular, wavefunctions
are singular at the singularity of the Schwarzschild(-like) black
hole. On the other hand, the geometry of the $SL(2,\IR)\over U(1)$
quotient CFT background is induced from the geometry of
$SL(2,\IR)$, and as a consequence includes regions beyond the
singularities. Moreover, T-duality interchanges the region outside
the horizon of the black hole with the one beyond the singularity
\cite{g,dvv} (for a review, see \cite{GPR}).

In string theory one is thus led to investigate the inclusion of
the regions beyond the singularity. A natural question to ask is:
What happens if one prepares a scattering wave beyond the
singularity? The answer to this question is given unambiguously by
doing the following. The wavefunctions in the quotient background
are obtained from gauge invariant vertex operators in the
$SL(2,\IR)$ CFT. As we shall describe later, one can obtain in
this way the behavior of scattering waves. In the uncharged 2-$d$
case it was found \cite{dvv} that low energy scattering waves,
obtained this way beyond the singularity, are fully reflected from
the singularity. We interpret this result as a confirmation that
the regions beyond the singularities of the uncharged black hole
can be ignored (as far as a low energy physics outside the black
hole is concerned).

In this note we consider the same issues for the two dimensional
{\it charged} black hole of \cite{ny}; related solutions appear in
\cite{ils,hh,hhs,GR,gq,quevedo}. It is very reminiscent of the
maximally extended Reissner-Nordstrom solution whose Penrose
diagram is shown in figure 2. In string theory such backgrounds
are obtained from a family of exact $\widehat{SL}(2,\IR)\times
U(1)\over U(1)$ quotient CFT sigma models, where
$\widehat{SL}(2,\IR)$ is the universal cover of $SL(2,\IR)$.
Again, in the 4-$d$ case each point in figure 2 is actually a two
sphere, while in the 2-$d$ case there is a non-trivial dilaton.
The singularity is time-like and may be avoided by free falling
neutral probes (see for instance section 3 in \cite{hh}). We
investigate if the singularity has other mild aspects in classical
string theory.

 \begin{figure}[h]
 \begin{center} \epsfxsize = 1.5in
\epsffile{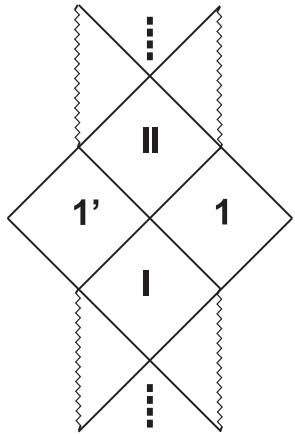} \caption{Penrose diagram of the
Reissner-Nordstrom black hole.} \label{fig 2} \end{center}
 \end{figure}

The geometry of the quotient CFT background is induced again from
the geometry of $\widehat{SL}(2,\IR)$ and, therefore, in string
theory we should consider including the regions beyond the
singularities. Moreover, scattering waves are uniquely determined
from vertex operators in the $SL(2,\IR)$ CFT. By following this
route we will show that scattering waves prepared beyond the
singularity are {\it not} fully reflected if the black hole is
charged. Part of the wavefunction is transmitted to regions
outside the black hole, in a manner consistent with unitarity. As
a consequence, initial data prepared beyond the singularities
affects observers outside the horizon of a charged black hole.

In section 2, we present a family of three dimensional backgrounds
obtained by gauging the WZW model of the four dimensional
$SL(2,\IR)\times U(1)$ group manifold by a family of non-compact
$U(1)$ subgroups. Then, by taking a small constant radius of the
$U(1)$ part, we obtain the two dimensional charged black hole via
the Kaluza-Klein (KK) mechanism. All two dimensional spaces have
asymptotically flat regions before event horizons and beyond inner
horizons, as well as regions between the event horizons and the
inner horizons. They admit singularities which are generated by
null gauge identifications and lie behind the inner horizons.
These backgrounds also have a non-trivial dilaton and a background
KK gauge field, and they can be obtained by $O(1,2)\subset O(2,2)$
rotations along the lines of \cite{GR} (for a review, see
\cite{GPR}).

In section 3, we discuss vertex operators in the quotient CFT. In
section 4, we present a class of solutions to the wave equation in
the ${SL(2,\IR)\times U(1)\over U(1)}$ CFT geometry which describe
scattering waves from beyond a singularity. One finds that such
scattering waves correspond to the vertex operators in the
quotient CFT. Moreover, we find that if the black hole is charged,
part of the wave which is coming from beyond the singularity leaks
into the region outside the horizon of the black hole. From the
point of view of the wavefunctions there is nothing special at the
location of the singularity. The results are discussed in section
5. In the appendix, we present the behavior of the wave scattered
from beyond the singularity in all the various regions of the
maximally extended charged black hole.

\section{The Geometry of ${SL(2,\IR)\times U(1)\over U(1)}$}
In this section we shall describe the geometry of the coset
${SL(2,\IR)\times U(1)\over U(1)}$ and its universal cover
${\widehat{SL}(2,\IR)\times U(1)\over U(1)}$. We construct a three
dimensional background by gauging \cite{brs} the WZW model of the
four dimensional $SL(2,\IR)\times U(1)$ group manifold by a
non-compact $U(1)$ subgroup. Let $(g,x)\in SL\(2,\IR\)\times
U\(1\)$ be a point on the product group manifold where $x\sim
x+2\pi L$, and let $k>0$ be the level of $SL(2,\IR)$ (such that
its signature is $(++-)$). The $U(1)$ gauge group acts as
 \eqn{gatr}{(g,x_L,x_R)\rightarrow (e^{\rho
\sigma_{3}/\sqrt{k} }g e^{\tau \sigma_{3}/\sqrt{k}},
 x_L+\rho',x_R+\tau')~,}
where $x_{L,R}$ are the left-handed and right-handed parts of $x$,
respectively. Since we gauge only a single $U(1)$ out of the two
right-handed $U(1)$ generators in \gatr, the two parameters
$(\tau,\tau')\equiv\underline\tau$ are not independent but rather
are constrained by
 \eqn{constr}{\underline\tau\equiv\tau\underline u~,}
where $\underline u$ is some unit real 2-vector. The left-handed
parameters $(\rho,\rho')\equiv\underline\rho$ in \gatr\ depend
linearly on the right-handed $\underline\tau$ parameters. For an
anomaly free gauging this dependence has to take the form
 \eqn{orth}{\underline{\rho} = R \underline\tau~, }
where the matrix $R$ is an $SO(2)$ matrix
\eqn{rmatrix}{R=\left(\matrix{\cos(\psi)&\sin(\psi)\cr
 -\sin(\psi)&\cos(\psi)}\right).}
The gauged action, as in \cite{egr}, is then defined by
 \eqn{gact}{S=S[e^{\hat\rho \sigma_{3}/\sqrt{k} }g
e^{\hat\tau \sigma_{3}/\sqrt{k}}]+ S'[x+\hat \rho'+\hat \tau']-
{1\over {2\pi}}\int d^2z (\partial\hat{ \underline {\rho}}
-R\partial\hat {\underline{ \tau}})^T (\bar{\partial}
\hat{\underline {\rho}} -R\bar
 {\partial} \hat{\underline {\tau}})~.}
Here, $S[g]$ is the WZW action,
 \eqn{wzg}{S[g]={k \over {4
\pi}}[\int_{\Sigma} Tr(g^{-1}\partial g g^{-1}\bar \partial g)-{1
 \over 3} \int_{B} Tr (g^{-1}dg)^3 ]~,}
where $\Sigma$ is the string's worldsheet and $B$ a 3-submanifold
of the group $SL(2,\IR)$ bounded by the image of $\Sigma$. $S'[x]$
is
 \eqn{uone}{S'[x]={1 \over {2 \pi}}\int_{\Sigma}\d x \bar \d x~.}
$\hat{\underline\rho}$ and $\hat{\underline\tau}$ are independent
fields subject to the constraints
 \eqn{taurhocnst}{\hat{\underline\tau}=(\hat{\underline\tau}^T
\underline u)\underline u\ ,\qquad \hat{\underline\rho}=
(\hat{\underline\rho}^T R\underline u )R\underline u~.}
 The action \gact\ is invariant under the gauge transformation
\gatr\ for the fields $g$ and $x$ together with the field
transformation
 \eqn{agatr}{\eqalign{&\hat{ \underline{\rho}}\rightarrow \hat
{\underline{\rho}}-\underline {\rho} \cr&\hat{\underline{
\tau}}\rightarrow \hat {\underline{\tau}}-\underline{\tau}}}
provided that the parameters $\underline{\rho}$ and
$\underline{\tau}$ satisfy the relation \orth. Using the
Polyakov-Wiegmann identity one sees that the action \gact\ depends
on $\hat{\underline{\rho}}$ and $\hat{\underline{\tau}}$ only
through the quantities
 \eqn{af}{\eqalign{&A ={\underline
u}^T\partial\hat{\underline{\tau}}\cr &\bar{A}=(R\underline
 u)^T\bar{\partial}\hat{\underline{\rho}}}}
The gauged action has then the form
 \eqn{act}{S= S[g]+S'[x]+{1 \over {2\pi}}\int
d^2z[ A\bar{\bf J}^T\underline u +\bar{A}{\bf J}^T R\underline u
 +2A\bar{A}\(R\underline u\)^T M \underline u]~.}
$A$ and $\bar A$ are holomorphic and anti-holomorphic gauge
fields. ${\bf J}^T$ and $\bar {\bf J}^T$ are the row vector of
currents,
 \eqn{cur}{\eqalign{{\bf J}^T&= (\sqrt{k}Tr[ \partial g
g^{-1}\sigma_3], 2\d x) \cr \bar{\bf J}^T&= (\sqrt{k}
 Tr[g^{-1} \bar{\partial} g\sigma_3], 2\bar{ \partial} x)}}
The $2\times 2$ matrix $M$ in \act\ is:
 \eqn{qfo}{M=\left(\matrix{{1\over 2}Tr[g^{-1}\sigma_3 g\sigma_3]
 &0\cr 0&1}\right) + R~. }
One can write the same action as a complete square
 \eqn{actsq}{\eqalign{S=& S[g]+S'[x]+
\cr +&{1 \over {2\pi}}\int d^2z\[\(\bar A+{\bar{\bf J}^T
\underline u \over 2\(R\underline u\)^T M \underline
u}\)2\(R\underline u\)^T M \underline u \(A+{{\bf J}^T R\underline
u\over 2\(R\underline u\)^T M \underline u}\)-{\(\bar{\bf J}^T
\underline u \)\({\bf J}^T  R\underline u\) \over 2\(R\underline
 u\)^T M \underline u}\]}}
After integrating out the fields $A$ and $\bar A$, to first order
in ${1\over k}$ the resulting action is~\footnote{In the
superconformal extension this background is claimed to be exact
\cite{bars}.}:
 \eqn{actaia}{S=S[g]+S'[x]-{1 \over {4\pi}}\int d^2z\[{\(\bar{\bf J}^T
\underline u \)\({\bf J}^T  R\underline u\) \over
 \(R\underline u\)^T M \underline u}\]}
and the dilaton, which is normalized such that $g_s=e^{\Phi}$,
becomes
 \eqn{dil}{\Phi = \Phi_0-{1\over 2}\log\(\(R\underline u\)^T M \underline u\)~.}
Now we have to choose a specific parameterization of $SL(2,\IR)$.
Since the gauge group acts on a group element $g\in SL(2,\IR)$ by
multiplication from the left and from the right in powers of
$e^{\sigma_3}$, it is convenient to represent the group elements
$g\in SL(2,\IR)$ by
 \eqn{grepr}{g(\alpha,\beta,\theta;\epsilon_1,\epsilon_2,\delta)=
e^{\alpha\sigma_3}(-1)^{\epsilon_1}(i\sigma_2)^{\epsilon_2}
 g_\delta(\theta)e^{\beta\sigma_3}=\(\matrix{a &b \cr c &d}\)~,}
where \eqn{where}{\epsilon_{1,2}=0,1~;\qquad\delta=I,1,1'~;\qquad
ad-bc=1~;\qquad a,b,c,d\in \IR~,}
 \eqn{gI}{g_I=\left(\matrix{ \cos\theta&-\sin\theta\cr
\sin\theta&\cos\theta\cr}\right);\qquad
 0\le\theta\le{\pi\over2}~,}
 \eqn{gone}{g_1=g_{1'}^{-1}= \left(\matrix{ \cosh\theta&\sinh\theta\cr
 \sinh\theta&\cosh\theta\cr}\right);\qquad 0\le\theta<\infty~,}
and $\sigma_i$ are the Pauli matrices:
 \eqn{sss}{\sigma_1=\left(\matrix{0&1\cr 1&0\cr}\right)~,\qquad
\sigma_2=\left(\matrix{0&-i\cr i&0\cr}\right)~,\qquad
\sigma_3=\left(\matrix{1&0\cr 0&-1\cr}\right)~.}

 \begin{figure}[h]
 \begin{center} \epsfxsize = 2.5in
\epsffile{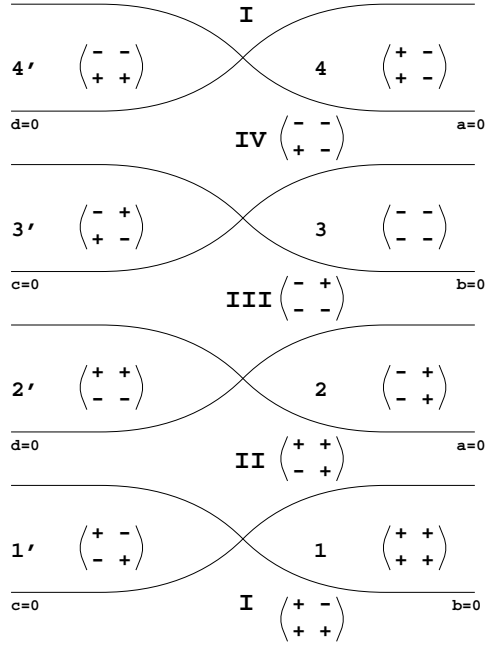} \caption{A two dimensional slice of
$SL(2,\IR)$.} \label{fig 3} \end{center}
 \end{figure}

\noindent This representation splits $SL(2,\IR)$ into twelve
regions\footnote{On the boundaries between the regions one has to
use a different representation, see \cite{egkr,vil}.} (see figure
3). Regions 1,1' and I are represented by
$\epsilon_1=\epsilon_2=0$ and $\delta=1,1',I$. The other nine
region are represented by different $\epsilon_{1,2}$. In different
regions the signs of the elements $a,b,c,d$ in $g$ \grepr\ are
different, as indicated in figure 3.

Define
 \eqn{defw}{W={\rm Tr} (\sigma_3 g\sigma_3 g^{-1})=2(2ad-1)=2(2bc+1)~,}
which is invariant under the gauge group action. Regions I,II,III
and IV (type B) have $|W|\le 2$. Regions 1,1',3 and 3' (type A)
have $W>2$. Regions 2,2',4 and 4' (type C) have $W<-2$ (for more
details see \cite{egkr}).

The gauge invariance of the action is fixed (for $\psi\neq\pi$) by
setting
 \eqn{gaugefix}{\alpha=-\beta\equiv{1\over 2}y~.}
After plugging \grepr\ in the gauge \gaugefix\ into \actaia, \dil\
one gets~\footnote{Equation \gaugefix\ does not fix the gauge at
$g_\delta=1$ in \grepr, but it turns out that the space is well
described by the coordinates \grepr\ (this is not the case for the
coordinates $u$ and $v$ that will be defined shortly).}
 \eqn{finalact}{\eqalign{S=&{1 \over 2\pi}\int_{\Sigma}\d x \bar \d
x+{k\over 2\pi}\int
d^2z\[-\d\theta_B\bar\d\theta_B+\sin^2(\theta_B)\d y\bar\d y\]+
\cr +&{1\over \pi}\int d^2z{\(\sqrt{k}\sin^2(\theta_B){\underline
u}_1\bar\d y-{\underline u}_2\bar\d x\)
\(\sqrt{k}\sin^2(\theta_B)(R\underline u)_1\d y+(R\underline u
)_2\d x\) \over \(R\underline u\)^T M \underline u} }}
\eqn{finaldil}{\Phi = \Phi_0-{1\over 2}\log\(\(R\underline u\)^T M
 \underline u \)~,}
where $|W|\le 2$. In the regions where $W>2$, $\theta_B$ in
\finalact, \finaldil\ should be replaced by $i\theta_A$. In the
regions with $W<-2$, substitute $i\theta_C$ for $\theta_B -
{\pi\over 2}$: \eqn{wwww}{\matrix{ B&\qquad |W|\le 2~,&\qquad
I,II,III,IV~:&\qquad \theta_B\cr A&\qquad W>2~,&\qquad
1,1',3,3'~:&\qquad \theta_B\rightarrow i\theta_A\cr C&\qquad
W<-2~,&\qquad 2,2',4,4'~:&\qquad \theta_B\rightarrow
i\theta_C+{\pi\over 2}~. }} If we take the vector
 \eqn{uchoice}{{\underline u}^T= (1,0)~,}
then $G_{xx}$ is constant~\footnote{Actually, $G_{xx}=const$ iff
$(G+B)_{yx}=0$ and, therefore, in this case the ${SL(2)\times
U(1)\over U(1)}$ background can be used in the heterotic string.}
and after rescaling $x \rightarrow \sqrt{k}x$ the action and the
dilaton become:
 \eqn{finalactu}{\eqalign{S=&{k \over
2\pi}\int_{\Sigma}\d x \bar \d x+{k\over 2\pi}\int d^2z \[-\d
\theta_B\bar \d \theta_B+\sin^2(\theta_B)\d y\bar\d y \]+ \cr
+&{k\over \pi}\int d^2z{\sin^2(\theta_B)\bar\d y
\(\sin^2(\theta_B)\cos(\psi)\d y-\sin(\psi)\d x\) \over
1+\cos(\psi)\cos(2\theta_B)}= \cr =&{k\over 2\pi}\int
d^2z\[-\d\theta_B\bar\d\theta_B+ {\d y\bar\d y -2p\bar\d y \d x
\over \cot^2(\theta_B)+p^2} + \d x \bar \d x\] }}
 \eqn{finaldilu}{\Phi = \tilde{\Phi}_0-{1\over
2} \log\(\cos^2(\theta_B)+p^2\sin^2(\theta_B)\)~, \qquad
\tilde{\Phi}_0\equiv\Phi_0+\half\log\({1+p^2\over 2}\)~,} where
\eqn{ppp}{p\equiv\tan({\psi\over 2})~.} Again, in regions A,C make
the appropriate replacement for $\theta_B$ (see eq. \wwww). For
large $k$ and for small radius of the circle parameterized by $x$,
the action \finalactu\ describes a two dimensional space-time
parameterized by $(\theta_i,y)$, $i=A,B,C$. The three dimensional
metric and antisymmetric background read from \finalact\, can be
reduced to two dimensions via the Kaluza-Klein mechanism. The term
proportional to $\d x \bar{\d} y$ gives rise in two dimensions to
a $U(1)$ gauge field whose charge is the momentum as well as the
winding along the $x$ circle. The two dimensional metric and
background gauge field take the form
 \eqn{twodimbg}{\eqalign{{1\over
k}ds^2&=-d\theta_B^2+{\cot^2(\theta_B) \over \(\cot^2(\theta_B)
+p^2\)^2}dy^2 \cr A_y&={\sqrt{k}p\over
 \cot^2(\theta_B)+p^2} }}
In regions A,C make the appropriate replacement for $\theta_B$
(see eq. \wwww). This is the two dimensional charged black hole
\cite{ny,GR}. To obtain the usual metric of the charged black hole
(in Schwarzschild-like coordinates) do the following coordinate
transformation: Rescale $y\rightarrow \(1-p^2\)t$, and define the
coordinate $r$ to be a linear function of the dilaton (in regions
of type A), given by
 \eqn{rcoor}{r={1\over 2}\log\(\sh^2(\theta_A)+{1\over
 1-p^2}\)~.}
The metric \twodimbg\ becomes
 \eqn{tdcbhmet}{{1\over k}ds^2= f(r )^{-1} dr^2 -f(r)dt^2~,}
where
 \eqn{fqm}{\eqalign{f(r)&=1-2me^{-2r}+q^2e^{-4r}\cr
 2m&={1+p^2 \over 1-p^2}~,\qquad q={p\over  1-p^2}~.}}
The background gauge field and dilaton are given by:
 \eqn{bgf_d}{ A_t(r)=\sqrt{k}\(qe^{-2r}-p\),\quad
 \Phi(r)=\tilde\Phi_0-{1\over 4}\log(1-p^2)^2-r\equiv\phi_0-r~.}
This metric describes a two dimensional charged black hole with
mass $M$ and charge $Q$ \cite{ny}:
 \eqn{mq}{M=me^{-2\phi_0}~,\qquad Q=qe^{-2\phi_0}~.}

\begin{figure}[h] \begin{center} \epsfxsize = 4in
\epsffile{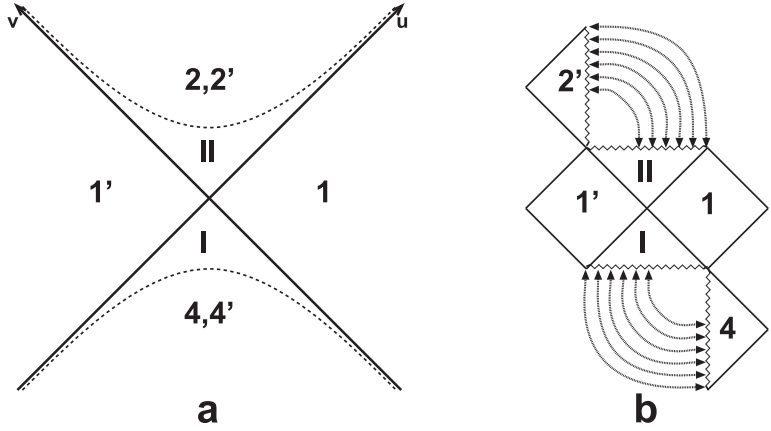} \caption{{\bf a}: The 2-dimensional black
hole ($\psi=0$); the solid lines are horizons and the dashed lines
are curvature singularities. {\bf b}: The Penrose diagram of the
2-dimensional black hole. The singularity is space-like when
viewed from regions I,II, and time-like from 2',4.} \label{fig 4}
 \end{center}
\end{figure}

\noindent In the ``Kruskal" coordinates~\footnote{The coordinates
$u$ and $v$ cover all of ${PSL(2)\over U(1)}$, and only half of
${SL(2)\over U(1)}$ (or the \Poincare patch of the universal
cover).} (for $|W|\le 2$)
 \eqn{uvcord}{u=\sin(\theta_B)e^y~, \quad v=\sin(\theta_B)e^{-y}}
the metric, dilaton and gauge field \finaldilu\ - \twodimbg\
are:~\footnote{For $\psi\neq 0$, $(\theta_B,y)$ are not good
coordinates at the points $\theta_B=\pm {\pi\over 2}$ (like the
origin of $\IR^2$ in polar coordinates). As a result, the
parametrization of these points is degenerated. The coordinate
transformation \uvcord\ is singular at $\theta_B=\pm {\pi\over 2}$
and, as a result, what looks like light-like lines at $uv=1$ are
actually the points $\theta_B=\pm{\pi\over 2}$.}
 \eqn{twodimbguv}{\eqalign{{1\over k}ds^2&={v^2du^2 +u^2dv^2\over
4uv}\( {1-uv\over \(1-(1-p^2)uv\)^2}-{1\over 1-uv}\) \cr
&-{dudv\over 2}\( {1-uv\over \(1-(1-p^2)uv\)^2}+{1\over 1-uv}\)}}
 \eqn{uvdilaton}{\Phi=\tilde\Phi_0-{1\over 4}\log \(1-(1-p^2)uv\)^2}
 \eqn{uvA}{A_u={\sqrt{k}\over2}{pv\over 1-\(1-p\)uv}~,\qquad
 A_v=-{\sqrt{k}\over2} {pu \over 1-\(1-p\)uv}~.}
For the degenerate case $\psi=0$ ($p=0$), the metric, dilaton and
background gauge field are:
 \eqn{twodbh}{\eqalign{ds^2&=-k{dudv\over
 1-uv} \cr \Phi&=\tilde\Phi_0-{1\over 4}\log (1-uv)^2} \qquad A_u=A_v=0}
This is the two dimensional Lorentzian black hole background. This
space is plotted in figure 4. The past and future horizons are
located at the $uv=0$ lines while the singularities are located at
the $uv=1$ lines. Regions of type A,C are static and approach flat
space at infinity. These are the regions before (after) the future
(past) horizons at $uv<0$ and beyond the singularities at $uv>1$.
Regions of type B are stretched between the horizons and the
singularities at $0\leq uv\leq 1$.

\begin{figure}[h] \begin{center} \epsfxsize = 4in
\epsffile{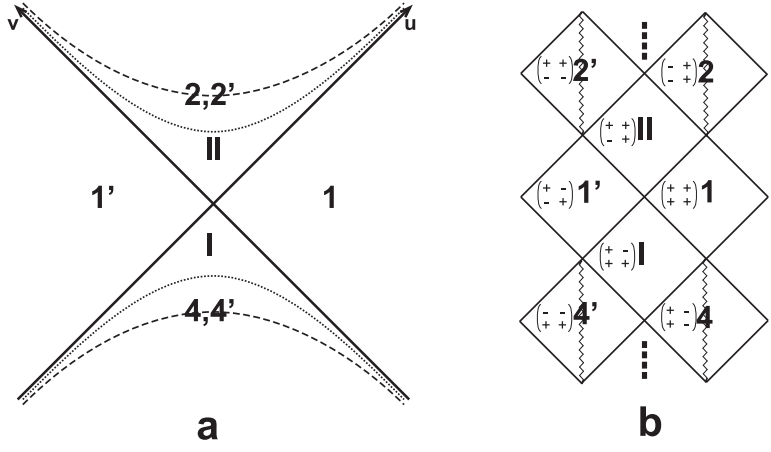} \caption{{\bf a}: The 2-dimensional charged
black hole ($\psi\neq 0$); the dashed lines are curvature
singularities and the doted lines (which are actually points) are
horizons. {\bf b}: The Penrose diagram of the 2-dimensional
charged black hole.} \label{fig 5}
 \end{center}
\end{figure}

When we charge the black hole ($\psi,p\neq 0$), the curvature
singularities move into regions of type C ($uv={1\over 1-p^2}>1$),
as indicated in figure 5. The lines $uv=1$ turn into inner
horizons. For $\psi=0$ the source of the singularity is a fixed
line of the gauge transformation and for $\psi\neq 0$ the source
is a null gauge orbit (see the appendix in \cite{egr}). Finally,
in the universal cover the geometry repeats itself.

\section{Vertex Operators in ${SL(2,\IR)\times U(1)\over U(1)}$}

Next, we would like to investigate the behavior of wavefunctions
in the 2-$d$ charged black hole geometry. However, we do not allow
arbitrary continuations of independent solutions across the
singularities. Instead, we are guided by the structure of the
vertex operators in the ${SL(2,\IR)\times U(1)\over U(1)}$ CFT.
Hence, in this section, we first discuss the (low lying) vertex
operators in ${SL(2,\IR)\times U(1)\over U(1)}$. Vertex operators
in a coset CFT are obtained by imposing the gauge conditions on
those in the underlying WZW model. Thus, we begin by inspecting
the low lying vertex operators in the $SL(2,\IR)$ CFT, or its
universal cover $\widehat{SL}(2,\IR)\equiv AdS_3$.

We are interested in the Lorentzian case, but since with a
Euclidean worldsheet the Euclidean $AdS_3 (\equiv H_3^+)$ is
better behaved, we first consider Euclidean  $AdS_3$. The elements
of $H_3^+$ can be parametrized by the $2\times 2$ matrices
\cite{tesch}: \eqn{ugg}{g=\(\matrix{a&b\cr c&d}\)= \(\matrix{
e^\phi\gamma & -e^{-\phi}-e^\phi\gamma
 \bar \gamma \cr e^\phi & -e^\phi\bar \gamma }\)=
\(\matrix{e^{-\phi}+e^\phi\gamma\bar \gamma & e^\phi\gamma\cr
e^\phi\bar \gamma & e^\phi}\)(-i\sigma_2) ~, } where $\phi\in
\IR$, $\gamma\in\IC$, and  $\bar\gamma=(\gamma)^*$. Low lying
primary vertex operators are eigenfunctions of the Laplacian on
$H^+_3$ with eigenvalues $-h(h-1)$. They are given by
\cite{tesch}:
 \eqn{fheig}{\Phi_h(x,\bar x;\phi,\gamma,\bar\gamma)=
 {2h-1\over \pi}\left(|\gamma-x|^2e^\phi+e^{-\phi}\right)^{-2h}~.}
The complex conjugate parameters $x, \bar x$ are interpreted as
the coordinates of the Euclidean $CFT_2$ dual to $AdS_3$.

Alternatively, we can write the vertex operators in ``momentum
space.'' We are interested in eigenfunctions of the transformation
$g\to e^{\alpha\sigma_3}ge^{\beta\sigma_3}$ with eigenvalues
$e^{2i(m\alpha+\bar m\beta)}$, $m,\bar m\in \IR$, and those are
given by~\footnote{These are the same as the operators usually
considered in CFT on $AdS_3$, but with $m$ replaced by $im$ and
$\bar m$ by $-i\bar m$ (see the discussion in section 2.2 of
\cite{egkr}).}:
 \eqn{khh}{K_{j;m,\bar m}(g)=\int d^2 x x^{j+im}\bar x^{j-i\bar m}
 \Phi_{j+1}(x,\bar x;\phi,\gamma,\bar\gamma)~.}
For future purposes we shall inspect the asymptotic behavior of
$K_{j;m,\bar m}$ at large $\phi$ and for $j+\half\in i\IR$. Using
the asymptotic behavior of $\Phi_h(x,\bar
x;\phi,\gamma,\bar\gamma)$ \cite{tesch} and eq. \khh, one finds:
 \eqn{katinfdis}{\eqalign{K_{j;m,\bar m}(g(\phi\rightarrow\infty))&\sim
\gamma^{im+j}\bar\gamma^{-i\bar m+j} e^{2j\phi}+ R(j;m,\bar m)
\gamma^{im-(j+1)}\bar\gamma^{-i\bar m-(j+1)}e^{-2(j+1)\phi} \cr
&\sim\left({-a\over d}\right)^{{i\over2}(m+\bar m)} \left({-b\over
 c}\right)^{{i\over2}(m-\bar m)}\[(-ad)^j+R(j;m,\bar m)(-ad)^{-(j+1)}\]~,}}
with
 \eqn{rjmbarm}{R(j;m,\bar m)={\Gamma(j+1+im)\Gamma(j+1+i\bar m)\Gamma(-2j-1)
 \over \Gamma(-j+im)\Gamma(-j+i\bar m)\Gamma(2j+1)}~.}
We now return to Lorentzian signature. The analytic continuation
from Euclidean to Lorentzian $AdS_3$ is obtained by taking
$\gamma$ and $\bar\gamma$ to be independent real parameters. In
this case $a,b,c,d$ in eq. \ugg\ are real and cover a Poincar\'e
patch of $AdS_3$. In the parametrization of eq. \grepr, large
$\phi$ corresponds to large $\theta$. For large $\theta$, near the
boundary of $SL(2,\IR)$, eq. \grepr\ reads:
 \eqn{asympg}{g(\theta\rightarrow \infty)\sim
e^\theta\(\matrix{\pm e^{\alpha+\beta} & \pm e^{\alpha-\beta} \cr
 \pm e^{\beta-\alpha} & \pm e^{-\alpha-\beta}}\)}
Hence, in the gauge \gaugefix, and for $j=-{1\over 2}+is$, the
asymptotic behavior of $K_{j;m,\bar m}$ is:
\eqn{asympofk}{K_{j;m,\bar m}\(g(\theta\rightarrow \infty)\)\sim
e^{-\theta+iy(m-\bar m)}\[e^{i2s\theta}+R(j;m,\bar
m)e^{-i2s\theta}\]~.} After adding the $U(1)$ contribution, the
vertex operator in $SL(2,\IR)\times U(1)$ becomes:
 \eqn{vvertex}{V^j_{m,k_L;\bar m, k_R}=K_{j;m,\bar m}
 (g)e^{i(k_Lx_L+k_R x_R)}~.}
Applying the gauge transformation \gatr\ to $(g,x)$, the operator
$V^j_{m,k_L;\bar m, k_R}$ gets multiplied by the phase
 \eqn{conchar}{e^{i({ m\over \sqrt{k}}\rho +k_L\rho'+
 {\bar m\over \sqrt{k}}\tau +k_R\tau')}~.}
In the ${SL(2,\IR)\times U(1)\over U(1)}$ coset, only those vertex
operators for which this phase equals to $1$ are allowed. Taking
\orth,\constr\ into account we get a constraint on the allowed
charges $(m,k_L;\bar m, k_R)$; in matrix notation it reads:
 \eqn{concharr}{\[({ m\over \sqrt{k}},k_L)R+
 ({\bar m\over \sqrt{k}},k_R)\]{\underline u}=0~,}
where $(k_L,k_R)\in\Gamma^{1,1}$ are quantized on the even
self-dual Narain lattice. For the choice \uchoice, the constraint
reads:
 \eqn{dispersion}{\bar m +m\cos (\psi)-\sqrt{k}k_L\sin(\psi)=0~.}
Finally, in string theory on ${SL(2,\IR)\times U(1)\over
U(1)}\times {\cal N}$, physical vertex operators are given by the
dressing of operators in the internal CFT ${\cal N}$ with
$V^j_{m,k_L;\bar m, k_R}$, such that the on-shell string
conditions are obtained. The precise details of the on-shell
conditions depend on the type of string theory considered.

\section{Scattering from Behind the Singularity}

In this section, we will show that vertex operators in the
${SL(2,\IR)\times U(1)\over U(1)}$ coset CFT describe scattering
waves in the 2-$d$ charged black hole geometry. Our focus will be
on waves which are incoming from a region behind a singularity,
say region 4 in figure 5 (though the discussion is easily extended
to scattering waves from asymptotically flat regions outside the
black hole, like region 1). For this purpose, we first discuss
independent wave solutions in $SL(2,\IR)$, and then consider
particular linear combinations which, when reduced to the coset,
describe scattering waves incoming entirely from region 4. These
scattering waves will turn out to correspond to the vertex
operators in the coset CFT discussed in section 3.

A general wave in $L_2\(\widehat{SL}(2,\IR)\)$ is given by a
linear combination of $\widehat{SL}(2,\IR)$ matrix elements in the
principal continuous and discrete series representations. Matrix
elements of $g$ in a representation with a Casimir $-j(j+1)$,
${\bf K}(j;g)$, are eigenfunctions of the Laplacian with
eigenvalue $-j(j+1)$. Principal continuous representations have
 \eqn{jcont}{j=-{1\over2}+is;\qquad s\in\IR~,}
and are further labelled by a phase $\exp(2\pi i\epsilon)$, where
$\epsilon=0$ in $PSL(2,\IR)$, $\epsilon=0,\half$ in $SL(2,\IR)$,
and $\epsilon\in (-\half,\half]$ for the universal cover
$\widehat{SL}(2,\IR)$. The phase $\exp(2\pi i\epsilon)$
corresponds to the representation of the center of the
corresponding group. The second class consists of the principal
discrete representations, characterized by real $j$, with
 \eqn{jdiscr}{j\in \IZ+\epsilon~.}
We will choose a basis of eigenvectors of the non-compact $U(1)$,
$g=\exp(\alpha\sigma_3)$. For unitary representations, the
corresponding eigenvalue is $\exp(2im\alpha)$, with $m\in\IR$. In
a given representation, $m$ can take any real value. Moreover, for
the continuous representations, there are two vectors with the
same value of $m$, which are distinguished by $\pm$.

Waves in the principal continuous series are
$\delta$-normalizable; these are the wavefunctions that we shall
study in this section. For the continuous representations in the
above basis, the non-vanishing matrix elements of $g$ \grepr\ are
given by:~\footnote{We will sometimes use the label $g$ both for
the $2\times 2$ matrices, as well as their representations.}
 \eqn{matelem}{\eqalign{&{\bf K}_{\pm\pm}(\lambda,\mu;j,\epsilon;g)
\equiv \langle j,\epsilon, m,\pm|g|j,\epsilon,\bar
m,\pm\rangle=\cr &e^{2i(m\alpha+\bar m\beta)}e^{2\pi i
\epsilon_1\epsilon} \langle j,\epsilon,
m,\pm|(i\sigma_2)^{\epsilon_2}
 g_\delta(\theta)|j,\epsilon,\bar{m},\pm\rangle~,\cr }}
where
 \eqn{lmmu}{\lambda\equiv -im-j;\;\; \mu\equiv -i\bar m-j~.}
These matrix elements appear in \cite{vil}\ (for the group
$SL(2,\IR)$).

Next we shall prepare a wave packet which, after reduction to the
coset, describes the scattering of a wave which is incoming from
behind the singularity of region 4. In region 4, where
$g(y,\theta_4)=e^{{y\over 2}\sigma_3}g_4(\theta_4) e^{-{y\over
2}\sigma_3}$ with $g_4=-i\sigma_2g_{1'}=g_1(-i\sigma_2)$ (see eqs.
\grepr\ -- \sss\ and figure 3):
\eqn{ggff}{g_4=e^{\theta_4\sigma_1}(-i\sigma_2)
=\(\matrix{\sinh\theta_4&-\cosh\theta_4\cr
\cosh\theta_4&-\sinh\theta_4}\)~,} the matrix elements in the
principle continuous series (with a given $j,\epsilon;m,\bar m$)
are:
 \eqn{kpm4}{\eqalign{{\bf K}_{+-}(\lambda,\mu;j,\epsilon;g_4)&=
{e^{2\pi i\epsilon}\over 2\pi i}
B(\lambda,-\lambda-2j){\ch^{\lambda-\mu}\theta_4 \over
\sh^{\lambda-\mu-2j}\theta_4}
F\(\lambda,-\mu-2j;-2j;-\sh^{-2}\theta_4\)~, \cr {\bf
K}_{-+}(\lambda,\mu;j,\epsilon;g_4)&= {1\over 2\pi i}
B(1-\mu,\mu+2j+1)\times \cr &\times{\ch^{\lambda-\mu}\theta_4
\over \sh^{\lambda-\mu+2j+2}\theta_4}
F\(1+\lambda+2j,1-\mu;2j+2;-\sh^{-2}\theta_4\)~,}}
 \eqn{kmm4}{\eqalign{&{\bf K}_{--}(\lambda,\mu;j,\epsilon;g_4)=
{{\bf K}_{+-}\over
B(\lambda,-\lambda-2j)}\[B(\lambda,2j+1)+e^{-2\pi
i\epsilon}B(-\lambda-2j,1+2j)\]+ \cr &+{{\bf K}_{-+}\over
B(1-\mu,\mu+2j+1)}\[e^{2\pi i\epsilon}B(1+\mu+2j,-2j-1)+
B(1-\mu,-2j-1)\] \cr &\equiv c_{+-}{\bf
K}_{+-}(\lambda,\mu;j,\epsilon;g_4)+
 c_{-+}{\bf K}_{-+}(\lambda,\mu;j,\epsilon;g_4)~,}}
 \eqn{kpp4}{{\bf K}_{++}=0~,}
where $B(a,b)$ is the Euler Beta function
 \eqn{bfff}{B(a,b)={\Gamma(a)\Gamma(b)\over\Gamma(a+b)}~,}
and $F(a,b;c;x)$ is the hypergeometric function ${}_2F_1$.

Wave functions (matrix elements) which are incoming only from
region 4 have to vanish on the border with region IV (one of the
$a=0$ lines in figure 3). For $a=0$ and $\psi\ne 0$ we can choose
the gauge
 \eqn{agauge}{g_\gamma=\(\begin{array}{cc}
 0 &-1\cr 1& -\gamma \end{array}\)}
The wave functions ${\bf K}_{+-},\ {\bf K}_{-+}$ on that line
(between region 4 and IV) are:
 \eqn{konline}{\eqalign{{\bf K}_{+-}(\lambda,\mu;j,\epsilon;g_\gamma)&=
{e^{2\pi i\epsilon}\over 2\pi i} B(\lambda,-\mu-\lambda-2j)
\gamma^{\mu+\lambda+2j}~, \cr {\bf
K}_{-+}(\lambda,\mu;j,\epsilon;g_\gamma)&={1\over 2\pi i}
 B(1+\mu+2j,-\lambda-\mu-2j)\gamma^{\mu+\lambda+2j}~.}}
Therefore, matrix elements in the continuous representations
corresponding to wave functions that are incoming from region 4
are given by:
 \eqn{www}{\eqalign{{\bf W}(\lambda,\mu;j,\epsilon;g)=
&{2\pi i\over B(\lambda,-\lambda-2j)}\[ e^{-2\pi i\epsilon}{\bf
K}_{+-}+{\sin(\pi(\mu+2j)) \over\sin(\pi\lambda)}{\bf K}_{-+}\]
\cr +&F(\lambda,\mu;j,\epsilon){\bf K}_{++} +
G(\lambda,\mu;j,\epsilon)\[{\bf K}_{--}-c_{+-}{\bf K}_{+-}-
 c_{-+}{\bf K}_{-+}\]~,}}
where $F(\lambda,\mu;j,\epsilon)$ and $G(\lambda,\mu;j,\epsilon)$
are arbitrary (normalizable) functions. A general wave which is
incoming from region 4 is given by a wave packet of ${\bf
W}(\lambda,\mu;j,\epsilon;g)$ in $\lambda,\mu,j$ and $\epsilon$.
Since we are interested only in the influence of a scattering from
region 4, where ${{\bf K}}_{++}={\bf K}_{--}-c_{+-}{\bf K}_{+-}-
c_{-+}{\bf K}_{-+}=0$ (although not in other regions), we will
concentrate on the case where $F=G=0$: When $F$ and/or $G$ are
non-zero, the second line in \www\ describes the physics of
wavefunctions entirely outside and independent of the data set in
region 4.

The asymptotic behavior of ${\bf K}_{+-}$ and ${\bf K}_{-+}$ in
region 4 is:
 \eqn{asymkinf}{\eqalign{{\bf K}_{+-}\(g_4(\theta_4\rightarrow\infty)\)=&
{e^{2\pi i\epsilon}\over 2\pi
i}B(\lambda,-\lambda-2j)e^{2j\theta_4+2i(m\alpha+\bar m \beta)}~,
\cr {\bf K}_{-+}\(g_4(\theta_4\rightarrow\infty)\) =&{1\over 2\pi
i}B(1-\mu,\mu+2j+1) e^{-2(j+1)\theta_4+2i(m\alpha+\bar m
\beta)}~,}} where $\theta_4\rightarrow\infty$ is the
asymptotically flat boundary of region 4. On the other hand,
\eqn{asymkzer}{\eqalign{&{\bf K}_{+-}\(g_4(\theta_4\rightarrow
0)\)= {e^{2\pi i\epsilon}\over 2\pi i}\({-b\over c}\)^{{i\over
2}(m-\bar m)}\times \cr
&\times\[B(-\mu-\lambda-2j,\lambda)d^{-i(m+\bar
m)}+B(-\lambda-2j,\lambda+\mu+2j)a^{i(m+\bar m)}\]~, \cr &{\bf
K}_{-+}\(g_4(\theta_4\rightarrow 0)\)={1\over 2\pi i}\({-b\over
c}\)^{{i\over 2}(m-\bar m)}\times \cr
&\times\[B(1+\mu+2j,-\mu-\lambda-2j)d^{-i(m+\bar m)}+B(1-\mu,
 \lambda+\mu+2j)a^{i(m+\bar m)}\]~,}}
where $\theta_4\rightarrow 0$ is the $a=d=0$ boundary between
region 4 and regions I and IV (see eq. \ggff\ and figures 3,5).
Equations \www,\asymkzer\ imply:
 \eqn{wasyzer}{\eqalign{&{\bf W}\(g_4(\theta_4\rightarrow 0)\)=
a^{i(m+\bar m)}\({-b\over c}\)^{{i\over 2}(m-\bar m)}\times\cr
&\times\[{B(-\lambda-2j,\lambda+\mu+2j)\over
B(\lambda,-\lambda-2j)}+{\sin(\pi(\mu+2j))
\over\sin(\pi\lambda)}{B(1-\mu,\lambda+\mu+2j)\over
 B(\lambda,-\lambda-2j)}\]~.}}
It is infinitely blue shifted as $a\rightarrow 0$: ${\bf W}\sim
a^{i(m+\bar m)}$, and hence any normalizable wave packet
constructed as a superposition of ${\bf W}$'s with different
values of $m$ and $\bar m$ vanishes at $a=0$, which is the
boundary between regions 4 and IV.

On the other hand, in the gauge \gaugefix\ ($\alpha=-\beta={y\over
2}$), eqs. \www,\asymkinf\ imply that the behavior of ${\bf W}$ in
the asymptotically flat region is:
 \eqn{wasyinf3}{{\bf W}\(g_4(\theta_4\rightarrow \infty)\)\sim
e^{-\theta_4+2i\omega t}\[e^{i2s\theta_4}+R(j;m,\bar
 m)e^{-i2s\theta_4}\]~,}
where
 \eqn{yw}{j=-\half+is,\qquad t=y~, \qquad \omega=\half (m -\bar m)~,}
and $R(j;m,\bar m)$ is given in \rjmbarm. This is precisely the
asymptotic behavior of the analytic continuation of the vertex
operators in the quotient CFT (see eq. \asympofk). Hence, we
conclude that ${\bf W}$ corresponds to the analytic continuation
of $V$ in the $SL(2)\times U(1)\over U(1)$ Euclidean
CFT~\footnote{The Euclidean quotient covers only a single
asymptotically flat region of the black hole. Here we have
analytically continued to region 4. The continuation to, say,
region 1 is done by taking $g\rightarrow gi\sigma_2$ in section 3
(thus interchanging, up to signs, the elements $a,b,c,d$ in eq.
\ugg), in which case we shall obtain the analog of the combination
${\bf U}$ of \cite{dvv,egkr}. The latter will describe a
scattering wave incoming from the asymptotically flat region
outside the black hole.}. The absolute value of $R$ is
 \eqn{absrrr2}{|R(j;m,\bar m)|^2={\ch\(\pi(2s-m-\bar m)\)+
\ch\(\pi(m-\bar m)\) \over\ch\(\pi(2s+m+\bar m)\)+
 \ch\(\pi(m-\bar{m})\)}~.}
For $\omega,s>0$, the combination ${\bf W}$ looks like a plane
wave which is incoming from the asymptotically flat boundary of
region 4, and scattered from the curved geometry. The damping
factor $e^{-\theta_4}$ is cancelled by a corresponding factor in
the $SL(2,\IR)$ measure. $R(j;m,\bar m)$ is the reflection
coefficient of this scattered wave. Indeed, $|R|\leq 1$ as it
should in a unitary theory. $R$ is also equal to the two point
function of $K_{j;m,\bar m}$ \khh, a primary field of
$SL(2)_L\times SL(2)_R$ with ``spin'' $j$ in the $SL(2)_k$ WZW
model in the large $k$ limit (see eq. (3.6) in
\cite{gk_comments}).

Note that for $s\neq 0$, $R(j;m,\bar m)$ is a phase if and only if
$\bar m=-m$. For neutral scattered particles ($(k_L,k_R)=0$) this
condition is satisfied if and only if the black hole is not
charged (see eq. \dispersion). We thus learn that an uncharged
wavefunction scattered from behind the singularity is fully
reflected if and only if the black hole is not charged. The fact
that the singularity of the neutral 2-$d$ black hole is a perfect
reflector was shown in \cite{dvv}.

The wave function ${\bf W}$ can be continued from region 4 to all
other regions of the maximally extended 2-$d$ charged black hole.
This is done by using the functions ${\bf K}$ in the various
regions. Some properties of the wave function ${\bf W}$ in the
extended black hole are presented in the appendix. In particular,
it is shown that the total incoming flux from all regions is equal
to the total outgoing flux, as expected in a unitary theory (the
unitarity being induced by that of $SL(2))$.

\section{Summary and Discussion}

To summarize, the results of section 4 indicate that in the two
dimensional charged black hole the regions beyond the
singularities should not be ignored, at least in classical string
theory. A scattering wave prepared behind the singularity is
transmitted ($|R|<1$), as long as the black hole is charged
($\psi\neq 0$ mod $\pi$). Moreover, the scattering wave $\bf W$ is
smooth at the singularity, though it is non-analytic (has an
infinitely blue shifted piece) at some of the horizons. Infinite
blue shifts at the horizons of charged black holes were investigated
both for the Reissner-Nordstrom black hole \cite{mzs,ch} and for
two dimensional charged black holes \cite{lmk,aa}. In both cases,
it leads to a large back reaction at the inner horizon and to a
potential instability.

We should emphasize that the analysis in this note is based on the
assumption that string perturbation theory is valid. Of course, a
large back reaction may invalidate this assumption. Indeed, as
mentioned above, since any perturbation in the geometry
\finalactu\ is infinitely blue shifted at the inner horizon, a
resulting singularity may be expected to form there
\cite{mzs,ch,lmk,aa}. This is one of the reasons why general
relativists do not consider seriously the regions behind the inner
horizon, where the singularity is located \cite{maldacena}.
However, the results in this work show that even if a singularity
is formed at the inner horizon, the region beyond it should not
necessarily be excluded. As we have seen, there is nothing
singular in the already existing singularity, at least as far as
low energy scattering from behind it is concerned.

The results in this work are supported by T-duality (for a review,
see \cite{GPR}). An axial-vector Abelian duality  corresponds to
taking $\psi\rightarrow \pi-\psi$. In particular, T-duality
interchanges the inner horizon at $uv=1$ with the event horizon at
$uv=0$, and takes the singularity at $uv>1$ to a line at $uv<0$,
which is a smooth line in the black hole geometry (similar to the
behavior of T-duality in the 3-$d$ black string background
considered in \cite{hh,quevedo}). Since nothing singular is
expected at the $uv<0$ regions -- neither for momentum modes nor
for ``windings'' -- such T-duality is compatible with our result
that nothing is singular about the singularity. In the uncharged
black hole $\psi=0$ the singularity coincides with the inner
horizon at $uv=1$, and T-duality interchanges the singularity with
the event horizon \cite{g,dvv}. As we charge the black hole,
$\psi$ increases, the singularity is split from the inner horizon
and moves towards $uv>1$ (see eqs. \twodimbguv,\uvdilaton). In the
extremal case $\psi={\pi\over 2}$ (in which case $M=Q$, see eqs.
\mq,\fqm,\ppp), the singularity in the $u,v$ coordinates
approaches $uv\rightarrow\infty$, and thus is ``removed'' (like in
\cite{egr}),~\footnote{Of course, in the Schwarzschild-like
coordinates $r,t$ \rcoor\ the two horizons coincide, instead.} and
as $\psi$ turns (formally) bigger then ${\pi\over 2}$ the
singularity ``re-appears'' in regions with $uv<0$. This is all
compatible, of course, with T-duality.

Finally, a very closely related two dimensional time-dependent
background in the presence of an Abelian gauge field is considered
in \cite{grs}. This 2-$d$ cosmological geometry is described
within a family of ${SL(2)_{k<0}\times U(1)\over U(1)\times \IZ}$
quotient CFTs. Hence, some of its aspects can be studied by
applying the same methods used in \cite{egkr,egr} and in this
note.

\bigskip\noindent {\bf Acknowledgements:} We are grateful to
S.~Elitzur and A. Konechny for collaborations and many
discussions. We thank O. Aharony, M. Berkooz, D. Gobonos, B. Kol,
D. Kutasov, Y. Oz, A. Pakman and M. Porrati  for discussions. This
work is supported in part by the BSF -- American-Israel
Bi-National Science Foundation, the Israel Academy of Sciences and
Humanities -- Centers of Excellence Program, the German-Israel
Bi-National Science Foundation, the European RTN network
HPRN-CT-2000-00122, and the Horwitz foundation (AS).

\begin{center} {\sc\Large Appendix} \end{center}

\appendix

\setcounter{equation}{0}

\section{Some Properties of the Scattering Wave W}

In this Appendix we give ${\bf W}$ in all the regions of a
Poincar\'e patch: 4',1,1',I and II (region 4 is studied in section
4). In particular, we present the asymptotic behavior of the
scattering wave {\bf W} in the asymptotically flat regions, and
check that the total incoming flux is equal to the total outgoing
flux.

Between asymptotically flat regions $a$ and $a'$, $a=1,..,4$, in
$SL(2,\IR)$ we have the relation
$g_{1'}(\theta)=g_{1}(-\theta)=-i\sigma_2g_{1}(\theta)i\sigma_2$
(see \gone), and between the ``intermediate'' regions $i=I,..,IV$
the relation is $g_I(-\theta)=\sigma_3g_I(\theta)\sigma_3$ (see
\gI). The relations in regions with $a>1$ and $i>I$ follow from
eqs. \grepr\ - \gone.~\footnote{In $\widehat{SL}(2,R)$ these
relations hold up to a jump between different copies (which is the
lift of $e^{2\pi i\sigma_2}$) and give an additional factor of
$e^{4\pi i\epsilon}$ to the matrix element.} In particular, these
relations give:
 \eqn{xxp}{{\bf K}_{\pm\mp}(\lambda,\mu;j,\epsilon;g_a)=
 {\bf K}_{\mp\pm}(\lambda,\mu;j,\epsilon;g_{a'})~, \qquad a=1,2,3,4~.}
We will use \xxp\ to read the behavior of ${\bf W}$ in region $a'$
from its behavior in region $a$.\\

\noindent{\bf In region 4' we have:}
 \eqn{wasyinf4'}{{\bf W}\(g_{4'}(\theta_{4'}\rightarrow \infty)\)=
e^{-\theta_{4'}-2i\omega t}\[A_{in\ 4'} e^{i2s\theta_{4'}} +
 A_{out\ 4'}e^{-i2s\theta_{4'}}\]~,}
where~\footnote{The interpretation of incoming is for incoming
positive energy particles (creation operators) or outgoing
negative energy particles (annihilation operators).}
 \eqn{ainout4'}{A_{in\ 4'}=e^{2\pi i\epsilon}{\sin(\pi(\mu+2j))
\over\sin(\pi\lambda)}~, \quad  A_{out\ 4'}= -e^{-2\pi i\epsilon}
{\Gamma(1-\mu)\Gamma(1+\mu+2j)\Gamma(-2j-1) \over
\Gamma(\lambda)\Gamma(-\lambda-2j)\Gamma(2j+1)}}
 \eqn{rt4'}{|A_{in\ 4'}|^2={\ch(\pi(\bar m-s))^2\over\ch(\pi(m+s))^2}~, \quad
|A_{out\ 4'}|^2={ \ch(\pi(m-s))\ch(\pi(m+s))\over
 \ch(\pi(\bar{m}-s))\ch(\pi(\bar m+s))}}\\

\noindent{\bf In region 1 we have:}
 \eqn{kpm1}{\eqalign{
{\bf K}_{-+}(\lambda,\mu;j,\epsilon;e^{\theta_1\sigma_1})= {1\over
2\pi i}\ch^{2j+\lambda+\mu}\theta_1 & \times\cr
\times[B(\lambda,1-\mu)\sh^{\lambda-\mu}\theta_1
&F\(\lambda,\lambda+2j+1;\lambda-\mu+1;-\sh^2\theta_1\)+ \cr
+e^{2\pi i\epsilon}B(-\lambda-2j,\mu+2j+1)
\sh^{\mu-\lambda}\theta_1
&F\(\mu,\mu+2j+1;\mu-\lambda+1;-\sh^2\theta_1\)]~, \cr
 {\bf K}_{+-}(\lambda,\mu;j,\epsilon;e^{\theta_1\sigma_1})&= 0~.}}
The asymptotic behavior in region 1 is
 \eqn{wasyinf1}{\eqalign{{\bf W}\(g_1(\theta_1\rightarrow \infty)\)=
&{2\pi i\sin(\pi(\mu+2j))\over \sin(\pi\lambda)
B(\lambda,-\lambda-2j)}{\bf K}_{-+} \(g_1(\theta_1\rightarrow
\infty)\) \cr =&e^{-\theta_1+2i\omega t}\[e^{i2s\theta_1}A_{in\
1}(j;m,\bar m) +
 e^{-i2s\theta_1}A_{out\ 1}(j;m,\bar m)\]~,}}
where
 \eqn{a1}{\eqalign{A_{in\ 1}(j;m,\bar m)&={\sin(\pi(\mu+2j))
 \over\sin(\pi\lambda)}{\Gamma(2j+1)\over
B(\lambda,-\lambda-2j)}\[{\Gamma(\lambda)\over
\Gamma(\lambda+2j+1)}+e^{2\pi i\epsilon} {\Gamma(-\lambda-2j)\over
\Gamma(1-\lambda)}\] \cr A_{out\ 1}(j;m,\bar
m)&={\sin(\pi(\mu+2j))
 \over\sin(\pi\lambda)}{\Gamma(-2j-1)\over B(\lambda,-\lambda-2j)}
\[{\Gamma(1-\mu)\over \Gamma(-\mu-2j)}+ e^{2\pi i\epsilon}
{\Gamma(\mu+2j+1)\over \Gamma(\mu)}\]}}
 \eqn{rt2'}{\eqalign{|A_{in\ 1}|^2&=
{\ch(\pi(\bar m-s))^2\over\ch(\pi(m+s))^2}{|\ch(\pi (m+s))+e^{2\pi
i\epsilon}\ch(\pi (m-s))|^2\over \sh^2(2\pi s)} \cr |A_{out\
1}|^2&={\ch(\pi(s-m))\ch(\pi(s-\bar m)) \over
\ch(\pi(s+m))\ch(\pi(s+\bar m))}{|\ch(\pi (\bar m+s))+e^{2\pi
i\epsilon}\ch(\pi (\bar m-s))|^2\over \sh^2(2\pi s)}}}

\noindent{\bf In region 1' we have:}
 \eqn{a2}{\eqalign{A_{in\ 1'}(j;m,\bar m)&=e^{-2\pi i\epsilon}
{\Gamma(2j+1)\over B(\lambda,-\lambda-2j)}
\[{\Gamma(\lambda)\over \Gamma(\lambda+2j+1)}+ e^{2\pi
i\epsilon}{\Gamma(-\lambda-2j)\over \Gamma(1-\lambda)}\] \cr
A_{out\ 1'}(j;m,\bar m)&= e^{-2\pi i\epsilon}{\Gamma(-2j-1)\over
B(\lambda,-\lambda-2j)}
\[{\Gamma(1-\mu)\over\Gamma(-\mu-2j)}+ e^{2\pi
i\epsilon}{\Gamma(\mu+2j+1)\over \Gamma(\mu)}\]}}
 \eqn{rt2}{\eqalign{|A_{in\ 1'}|^2&={|\ch(\pi
(m+s))+e^{2\pi i\epsilon}\ch(\pi (m-s))|^2\over \sh^2(2\pi s)} \cr
|A_{out\ 1'}|^2&={\ch(\pi(s+m))\ch(\pi(s-m)) \over
\ch(\pi(s+\bar{m})) \ch(\pi(s-\bar m))} {|\ch(\pi (\bar
m+s))+e^{2\pi i\epsilon}\ch(\pi (\bar m-s))|^2\over \sh^2(2\pi
 s)}}}

\noindent{\bf In region I we have ($-{\pi\over 2}<\theta_{I}<0$):}
 \eqn{www1p}{\eqalign{
&{\bf W}(\lambda,\mu;j,\epsilon;e^{i\theta_{I}\sigma_2})= \cr
&{B(-\lambda-2j,\mu+2j+1)\over
B(\lambda,-\lambda-2j)} {\sin^{\mu-\lambda}\theta_{I}\over
\cos^{\mu-\lambda-2j}\theta_{I}}
 F\(-\lambda-2j,\mu;\mu-\lambda+1;-\tan^2\theta_{I}\)+ \cr
&+{\sin(\pi(\mu+2j))\over\sin(\pi\lambda)}{B(\lambda,1-\mu)\over
B(\lambda,-\lambda-2j)} {\sin^{\lambda-\mu}\theta_{I}\over
\cos^{\lambda-\mu-2j}\theta_{I}}
 F\(\lambda,-\mu-2j;\lambda-\mu+1;-\tan^2\theta_{I}\)}}

\noindent{\bf In region II we have ($0<\theta_{II}<{\pi\over
2}$):}
 \eqn{wwwII}{\eqalign{
&{\bf W}(\lambda,\mu;j,\epsilon;e^{i\theta_{II}\sigma_2})= \cr
&e^{-2\pi i\epsilon}{B(\lambda,1-\mu)\over B(\lambda,-\lambda-2j)}
{\sin^{\lambda-\mu}(-\theta_{II})\over
\cos^{\lambda-\mu-2j}\theta_{II}}
F\(\lambda,-\mu-2j;\lambda-\mu+1;-\tan^2\theta_{II}\)+ \cr
&+e^{2\pi i\epsilon}{\sin(\pi(\mu+2j))\over\sin(\pi\lambda)}
{B(-\lambda-2j,\mu+2j+1)\over B(\lambda,-\lambda-2j)}\times\cr
&\times {\sin^{\mu-\lambda}(-\theta_{II})\over
\cos^{\mu-\lambda-2j}\theta_{II}}
 F\(-\lambda-2j,\mu;\mu-\lambda+1;-\tan^2\theta_{II}\)}}

\noindent The behavior of {\bf W} in all other regions in
$\widehat{SL}(2,\IR)$ is obtained by the relation~\footnote{And
the phase $e^{4\pi i\epsilon}$ that the wavefunction acquires when
we go from one copy of $SL(2,\IR)$ to the next one in the
universal cover.}
 \eqn{pmg}{{\bf K}_{\pm\pm}(\lambda,\mu;j,\epsilon;-g)=
 e^{2\pi i\epsilon}{\bf K}_{\pm\pm}(\lambda,\mu;j,\epsilon;g)~.}
Note that, for every $m,\bar m,s$ and $\epsilon$, the total
incoming flux is equal to the total outgoing flux, namely:
 \eqn{total}{\(1+|A_{in\ 4'}|^2+|A_{in\ 1}|^2+|A_{in\
 1'}|^2\)-\(|R|^2+|A_{out\ 4'}|^2+|A_{out\ 1}|^2+|A_{out\
 1'}|^2\)=0~,}
where we have normalized $A_{in\ 4}=1$, hence $R\equiv A_{out\
4}$. Finally, for $\bar m=-m$ ($\psi=0$) we have $|A_{in\
4}|^2=|A_{in\ 4'}|^2=|A_{out\ 4'}|^2=|R|^2=1$, and for
$\epsilon=0$, $\bar m=m$ ($\psi=\pi$) ${\bf W}$ vanish in region
II.

 \end{document}